\documentclass[a4paper]{article}
\usepackage{interspeech2009,amssymb,amsmath,epsfig}
\usepackage{graphicx}
\ninept
\def\reg{{\rm\ooalign{\hfil
     \raise.07ex\hbox{\scriptsize R}\hfil\crcr\mathhexbox20D}}}

\title{Glottal Closure and Opening Instant Detection from Speech Signals}

\makeatletter
\def\name#1{\gdef\@name{#1\\}}
\makeatother
\name{{\em Thomas Drugman, Thierry Dutoit}}

\address{TCTS Lab, Facult\'e Polytechnique de Mons - 31, Boulevard Dolez, 7000, Mons, Belgium \\}

\begin{document}
\maketitle
\begin{abstract}
This paper proposes a new procedure to detect Glottal Closure and Opening Instants (GCIs and GOIs) directly from speech waveforms. The procedure is divided into two successive steps. First a mean-based signal is computed, and intervals where speech events are expected to occur are extracted from it. Secondly, at each interval a precise position of the speech event is assigned by locating a discontinuity in the Linear Prediction residual. The proposed method is compared to the DYPSA algorithm on the CMU ARCTIC database. A significant improvement as well as a better noise robustness are reported. Besides, results of GOI identification accuracy are promising for the glottal source characterization.
\end{abstract}


\section{Introduction}\label{sec:intro}

In speech processing, Glottal Closure Instants (GCIs) are referred to the instances of significant excitation of the vocal tract. These particular time events correspond to the moments of high energy in the glottal signal during voiced speech.  Knowing the GCI location is of particular importance in speech processing.

For speech analysis, closed-phase LP autoregressive analysis techniques have been developed for better estimating the prediction coefficients, which results in a better estimation of the vocal tract resonances \cite{Childers}. These techniques explicitly require the determination of GCIs. A wide range of applications also implicitly assume that these instants are located. In concatenative speech synthesis, it is well known that some knowledge of a reference instant is necessary to eliminate concatenation discontinuities. This motivated the use of GCIs in the famous TD-PSOLA algorithm \cite{PSOLA} or as a means to remove phase mismatches \cite{Stylianou}. GCI has also been used for voice transformation \cite{VoiceConversion}, voice quality enhancement \cite{Naylor}, speaker identification \cite{SpeakerIdentification}, glottal source estimation \cite{ZZT}, or speech coding and transmission \cite{PSCELP}.

Many methods have been proposed to locate the GCIs directly from speech waveforms. The earliest attempts relied on the determinant of the autocovariance matrix \cite{Strube}. A study of the use of the Linear Prediction (LP) residual was investigated in \cite{Yegna}. Indeed, as GCIs correspond to instants of significant excitation, it is assumed that a large value in the LP residual is informative about the GCI location. In \cite{Frobenius}, GCIs were determined as the maxima of the Frobenius norm. An approach based on a weighted nonlinear prediction was proposed in \cite{Schnell}. In \cite{wavelet}, an algorithm based on a wavelet decomposition was considered. Some techniques also exploit the phase properties due to the impulse-like nature at the GCI by computing a group delay function \cite{GD}. The DYPSA algorithm, presented in \cite{Dypsa}, estimates GCI candidates using the projected phase-slope and employs dynamic programming to retain the most likely ones. In \cite{Kahawara}, GCIs are located by the center-of-gravity based signal and then refined by using minimum-phase group delay functions derived from the amplitude spectra. More recently, authors in \cite{Murty} proposed to detect discontinuities in frequency by confining the analysis around a single frequency. In this latter work, GCIs correspond to the positive zero-crossings of a filtered signal obtained by successive integrations of the speech waveform and followed by a mean removal operation. Comparative studies of the most popular approaches were led in \cite{Dypsa} and \cite{Murty}. It was shown that the DYPSA algorithm and the technique proposed in \cite{Murty} clearly outperformed other state-of-the-art methods.

On the other hand, very few works addressed the determination of Glottal Opening Instants (GOIs) from speech signals. Indeed, as the energy of excitation at GOIs is known to be weaker and more dispersed (resulting in more regular behaviour) than at GCIs \cite{Dypsa}, their automatic location remains a challeging problem. A method based on a multiscale product of wavelet transforms was proposed in \cite{Tunisian}, but no quantitative results were given.

This paper proposes a simple procedure to detect GCIs and GOIs from speech waveforms. The procedure is divided into two steps. First, an initial estimate of the GCI location is computed from a mean-based signal. This latter is obtained by calculating the mean of sliding windowed speech segments. This first estimation gives short intervals where GCIs are expected to occur. The second step aims at refining the GCI location by finding, for each interval, the largest LP residual value, which is assumed to correspond to the strongest impulse in the excitation signal.

The paper is structured as follows. Our proposed method is fully described in Section \ref{sec:method}. In Section \ref{sec:results}, we present our results obtained on the CMU ACRTIC database \cite{CMU}. As the performance of our technique depends on the window length used for computing the mean-based signal, the impact of this parameter is first discussed. We then compare our method with the DYPSA algorithm \cite{Dypsa} according to their GCI detection performance. The accuracy we obtained on GOI determination is also presented. Besides the noise robustness of both techniques is analyzed. Finally we conclude in Section \ref{sec:conclu}.

\section{Proposed method}\label{sec:method}

The proposed method consists of two successive steps. During the first step (Section \ref{ssec:mean}), a mean-based signal is computed, allowing the determination of short intervals where GCIs and GOIs are expected to occur. As for the second step (Section \ref{ssec:refinement}), it consits of a refinement of the accurate locations from the LP residual signal. 

\subsection{Interval determination from a mean-based signal}\label{ssec:mean}

In \cite{Murty}, authors argue that a discontinuity in the excitation is reflected over the whole spectral band, including the zero frequency. For this, they use the output of 0-Hz resonators to locate GCIs. Inspired from this observation, we focus our analysis on a mean-based signal. If $s(n)$ denotes the speech waveform, the mean-based signal $y(n)$ is computed as:

\begin{equation}\label{eq:Mean}
y(n)=\frac{1}{2N+1}\sum_{m=-N}^N{w(m)s(n+m)}	
\end{equation}

where $w(m)$ is a windowing function of length $2N+1$. In our experiments we used a Blackman window whose length is chosen as explained in Section \ref{ssec:WinLength}.

\begin{figure}[!ht]
  \centering
  \includegraphics[width=0.5\textwidth]{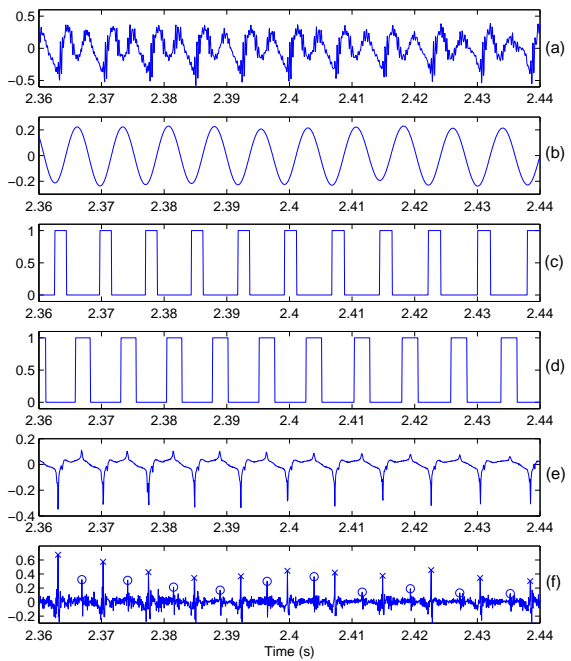}
  \caption{Example of GCI and GOI extraction on a voiced segment: (a) the speech signal, (b) its corresponding mean-based signal, (c) interval of GCI presence derived from the mean-based signal (between the minimum and the following positive zero-crossing), (d) interval of GOI presence derived from the mean-based signal (between the maximum and the following negative zero-crossing, with a margin of 0.25 ms), (e) aligned differenced electroglottograph, (f) the LP residual with the detected GCIs (x) and GOIs (o).}
  \label{fig:TotalFig}
\end{figure}

Figures \ref{fig:TotalFig}(a) and \ref{fig:TotalFig}(b) show an example of a voiced speech segment together with its corresponding mean-based signal. This latter presents the important property to evolve at the local pitch rhythm. However this signal in itself is not sufficient for accurately locating the GCIs. Indeed, we reported through our observations that a GCI occurs at a non-constant position between the minimum and the following positive zero-crossing of the mean-based signal. For this, we define intervals where the precise location of the GCI is expected to lie. In the same way, we observed that the GOI position falls within an interval defined by the maximum and the following negative zero-crossing of the mean-based signal. To ensure that the previous interval contains the real GOI, a margin of 0.25 ms is added at both sides of it. In addition, to avoid a possible irrelevant drift in the mean-based signal, previous zero-crossings are replaced by the midpoints between two successive extrema. Figures \ref{fig:TotalFig}(c) and \ref{fig:TotalFig}(d) exhibit such intervals extracted from the mean-based signal of Fig. \ref{fig:TotalFig}(b) respectively for GCIs and GOIs.

\subsection{GCI and GOI location refinement from the LP residual}\label{ssec:refinement}
Intervals obtained in the previous Section give "fuzzy" short regions where particular events (GCI or GOI) should happen. The goal of the current step is to associate an accurate location of an event within an interval. For this, we rely on the Linear Prediction (LP) residual. Indeed, after removing an approximation of the vocal tract response, one can expect that significant impulses in the excitation signal will be reflected in the LP residual. We can consequently assume that the event location corresponds to the strongest peak of the LP residual within the interval. Figures \ref{fig:TotalFig}(e) and \ref{fig:TotalFig}(f) show the time-aligned differenced electroglottograph (EGG) and the LP residual. Combining the intervals extracted from the mean-based signal with a peak picking method on the LP residual allows to accurately and unambiguously detect both GCIs and GOIs. Nervertheless while the impulse at the GCI significantly emerges from its neighborhood, the behaviour at the GOI is more regular since the excitation presents a discontinuity more spread out and with a weaker strength. As a consequence, obtaining for GOIs an identification accuracy  comparable to what can be achieved for GCIs remains a challenging problem (cf Section \ref{ssec:accuracy}).

\section{Results}\label{sec:results}

The experiments presented in this Section were achieved on the CMU ARCTIC database (publicly available in \cite{CMU}) containing 3 speakers: BDL (US male), JMK (Canadian male) and SLT (US female). The database consists of 1132 phonetically balanced utterances for each speaker (about 50 min), giving a total duration of around 2h40min. We compare our proposed method with the DYPSA algorithm \cite{Dypsa} whose implementation can be found in \cite{DypsaWeb}. Both techniques are applied on 16 kHz speech waveforms and EGG signals are used as a reference. Note that EGGs were time-aligned to compensate the delay between the laryngograph and the microphone. A 24-th order LP analysis was performed on 25ms long Hanning-windowed frames, shifted every 5 ms, and the LP residual was obtained by inverse filtering. To assess the performance of the methods we employed the measures defined in \cite{Dypsa}, namely:
\begin{itemize}
\item the Identification Rate (IDR),
\item the Miss Rate (MR),
\item and the False Alarm Rate (FAR),
\end{itemize}
and two indicators characterizing the timing error probability density:
\begin{itemize}
\item the Identification Accuracy (IDA), i.e the standard deviation of the distribution,
\item the accuracy to $\pm$ 0.25 ms, i.e the rate of detections for which the timing error is smaller than this bound.
\end{itemize}

\subsection{Impact of the window length}\label{ssec:WinLength}
\begin{figure}[!ht]
  \centering
  \includegraphics[width=0.49\textwidth]{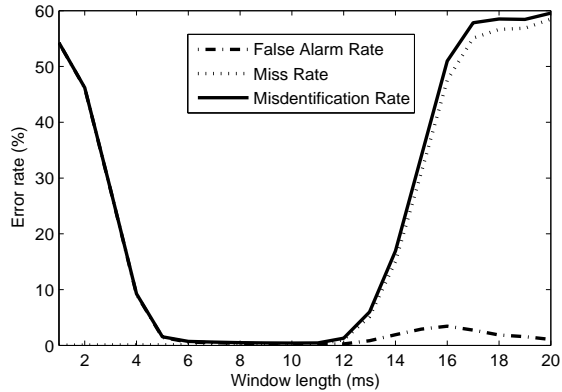}
  \caption{Effect of the window length on the misidentification rate for the speaker SLT, whose average pitch period is 5.7 ms.}
  \label{fig:Window}
\end{figure}

As explained in Section \ref{ssec:mean}, our method is controled by only one parameter (once the LP analysis is fixed): the window length used in Equation \ref{eq:Mean}. The influence of this parameter on the misidentification rate ($=1-IDR$) is illustrated in Figure \ref{fig:Window} for the female speaker SLT. Optimality is seen as a trade-off between two opposite effects. A too short window causes the appearance of spurious extrema in the mean-based signal, giving birth to false alarms. On the other hand, a too large window smooths it, affecting in this way the miss rate. However we clearly observed for the three speakers a valley between 1.5 and 2 times the average pitch period $T_{0,mean}$. Throughout the rest of this article we used a window whose length is 1.75$\cdot$$T_{0,mean}$. A pitch-dependent approach could also be envisaged but with the drawback of requiring a reliable pitch estimator.

\subsection{Identification performance}\label{ssec:accuracy}
\begin{figure}[!ht]
  \centering
  \includegraphics[width=0.49\textwidth]{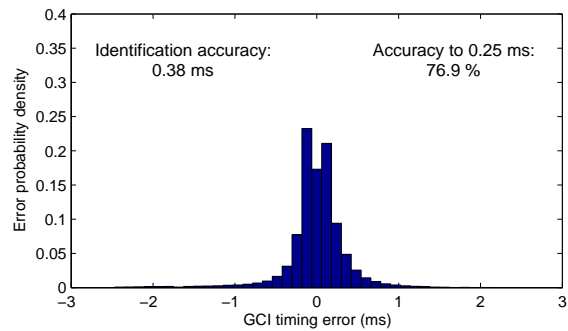}
  \caption{Histogram of the GCI timing error averaged over all speakers for the DYPSA algorithm.}
  \label{fig:DypsaHist}
\end{figure}

\begin{figure}[!ht]
  \centering
  \includegraphics[width=0.49\textwidth]{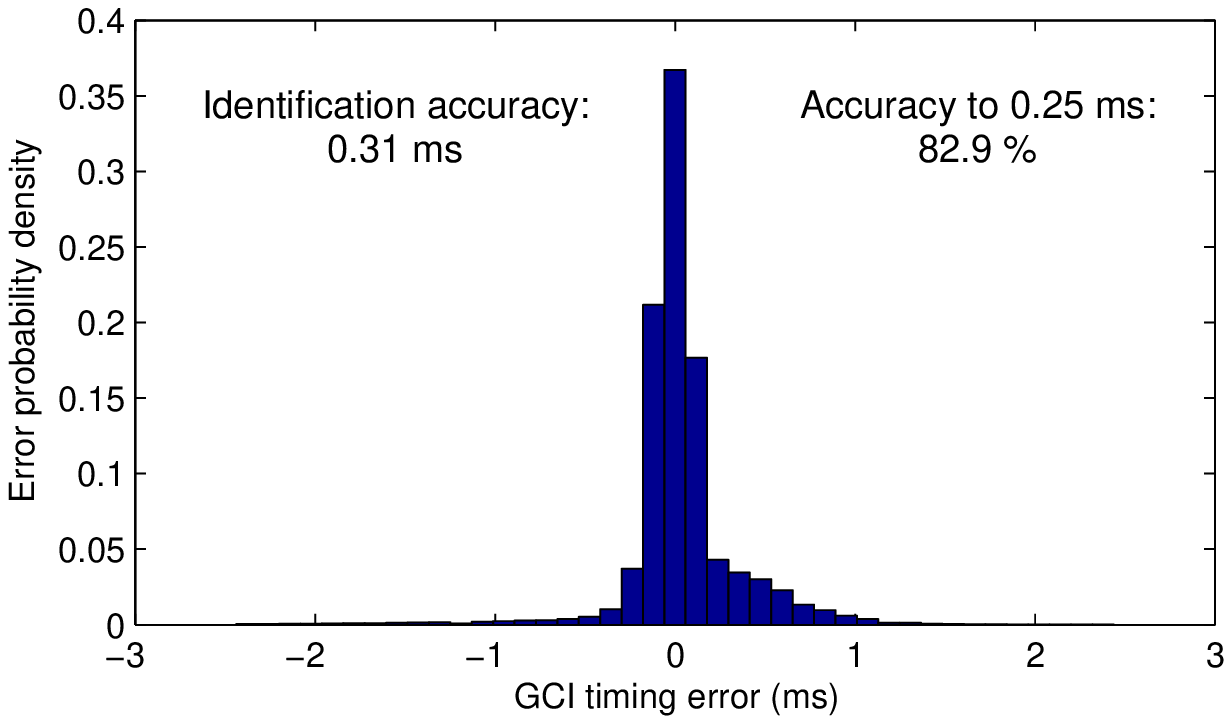}
  \caption{Histogram of the GCI timing error averaged over all speakers for the proposed method.}
  \label{fig:MeanHistGCI}
\end{figure}

\begin{figure}[!ht]
  \centering
  \includegraphics[width=0.49\textwidth]{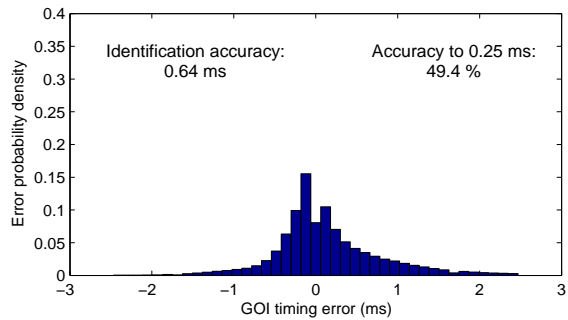}
  \caption{Histogram of the GOI timing error averaged over all speakers for the proposed method.}
  \label{fig:MeanHistGOI}
\end{figure}

Table \ref{tab:IDrate} details the identification efficiency for both DYPSA and proposed methods. A clear advantage can be noticed in favor of our technique over all rates and speakers. Since this performance is conditioned in our method by the mean-based signal, results are sensibly the same for GCI and GOI detection. On the opposite, error probability densities depend on the LP-based location refinement step and the accuracy consequently differs for GCIs and GOIs. Table \ref{tab:IDaccuracy} summarizes comparative accuracy results for the DYPSA algorithm and for our method employed for GCI as well as GOI detection. It can be noted that our proposed technique outperforms DYPSA except for speaker JMK whose results are almost similar. It also turns out that GOIs are less precisely located than GCIs, which was expected for the reasons underlined in Section \ref{ssec:refinement}. Nonetheless, despite these inherent difficulties, the proposed technique appears to give a rather efficient estimation of the GOI position. Leading to the same conclusions, figures \ref{fig:DypsaHist}, \ref{fig:MeanHistGCI} and \ref{fig:MeanHistGOI} depict the histograms averaged over all the speakers of the timing error made by the DYPSA algorithm on the GCI determination, and by the proposed method on both GCIs and GOIs respectively. Among others, it can be seen that our technique is more accurate than DYPSA and that 84\% of identified GOIs are located with an absolute error lower than 1 ms.

\begin{table}[!ht]
\centering
\begin{tabular}{c c c c c}
\hline \hline
Speaker & Method & IDR (\%) & MR (\%) & FAR (\%) \\
\hline \hline
BDL & Dypsa & 96.81 & 1.78 & 1.41 \\ 
BDL & Proposed & 98.89 & 0.61 & 0.50 \\ 
\hline
JMK & Dypsa & 98.17 & 1.50 & 0.33 \\ 
JMK & Proposed & 98.59 & 1.30 & 0.11 \\ 
\hline
SLT & Dypsa & 97.44 & 1.43 & 1.13 \\ 
SLT & Proposed & 99.34 & 0.17 & 0.49 \\ 
\hline
\end{tabular}
\caption{Comparative results in terms of Identification Rate (IDR), Miss Rate (MR) and False Alarm Rate (FAR).}
\label{tab:IDrate}
\end{table}

\begin{table}[!ht]
\centering
\begin{tabular}{c c c c c}
\hline \hline
Speaker & Method & Event & IDA (ms) & Accuracy to\\
  &   &   &   & $\pm$ 0.25 ms (\%) \\
\hline \hline
BDL & Dypsa & GCI & 0.34 & 81.7\\ 
BDL & Proposed & GCI & 0.25 & 88.8\\ 
BDL & Proposed & GOI & 0.49 & 65.2\\ 
\hline
JMK & Dypsa & GCI & 0.41 & 74.2\\ 
JMK & Proposed & GCI & 0.41 & 74.0\\ 
JMK & Proposed & GOI & 0.69 & 48.3\\ 
\hline
SLT & Dypsa & GCI & 0.38 & 75.1\\ 
SLT & Proposed & GCI & 0.27 & 83.5\\ 
SLT & Proposed & GOI & 0.63 & 41.2\\ 
\hline
\end{tabular}
\caption{Comparative results in terms of Identification Accuracy (IDA) and accuracy to $\pm$ 0.25 ms, characterizing the error probability densities.}
\label{tab:IDaccuracy}
\end{table}

\subsection{Noise robustness}
\begin{figure}[!ht]
  \centering
  \includegraphics[width=0.49\textwidth]{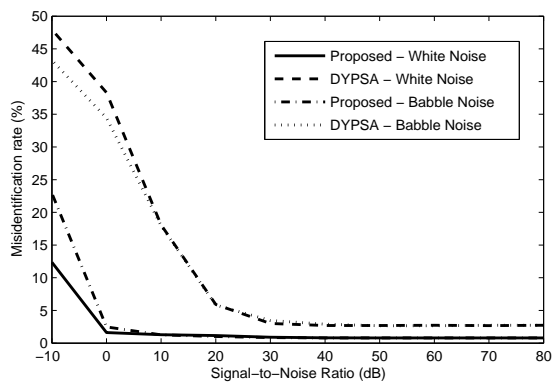}
  \caption{Comparison of the performance degradation with additive white and babble noises for the DYPSA and proposed methods.}
  \label{fig:Noise}
\end{figure}

Methods are here compared according to their noise robustness. For this, a white Gaussian noise and a babble noise (from a cafetaria environment) were added at different levels to the speech signals. The Signal-to-Noise Ratio (SNR) varies from -10 dB (extremely adverse conditions) to 80 dB (almost clean speech). Figure \ref{fig:Noise} reports the evolution of the misidentification rate with the noise level. Our technique remains almost insensitive up to 0 dB while DYPSA begins to degrade from 30 dB before being severly affected from 10 dB. This observation holds for both noise type.

\section{Conclusion}\label{sec:conclu}

This paper proposed a new procedure for detecting the GCIs and GOIs directly from speech signals. The procedure was divided into two successive steps. The first one computed a mean-based signal and extracted from it intervals where speech events were expected to occur. This step guaranteed good performance in terms of identification rate. The second one refined the location of the speech events within the intervals by inspecting the LP residual. As for it, this step ensured good performance in terms of identification accuracy. Our proposed method was compared to the DYPSA algorithm on the CMU ARCTIC database. Through our experiments, we reported a significant improvement in GCI detection efficiency as well as in noise robustness. In addition our method also allowed to determine the GOIs locations with an encouraging precision, although not yet comparable to what can be achieved for GCIs. As future work we plan to enhance the GOI locations by analyzing the open quotient trajectories. We also plan to investigate the characterization of the glottal source by combining the proposed method with other source-filter deconvolution approaches. 

\section{Acknowledgments}\label{sec:Acknowledgments}

Thomas Drugman is supported by the ``Fonds National de la Recherche
Scientifique'' (FNRS).

\eightpt
\bibliographystyle{IEEEtran}

\begin{thebibliography}{}
\providecommand{\url}[1]{#1}
\csname url@samestyle\endcsname
\providecommand{\newblock}{\relax}
\providecommand{\bibinfo}[2]{#2}
\providecommand{\BIBentrySTDinterwordspacing}{\spaceskip=0pt\relax}
\providecommand{\BIBentryALTinterwordstretchfactor}{4}
\providecommand{\BIBentryALTinterwordspacing}{\spaceskip=\fontdimen2\font plus
\BIBentryALTinterwordstretchfactor\fontdimen3\font minus
  \fontdimen4\font\relax}
\providecommand{\BIBforeignlanguage}[2]{{%
\expandafter\ifx\csname l@#1\endcsname\relax
\typeout{** WARNING: IEEEtran.bst: No hyphenation pattern has been}%
\typeout{** loaded for the language `#1'. Using the pattern for}%
\typeout{** the default language instead.}%
\else
\language=\csname l@#1\endcsname
\fi
#2}}
\providecommand{\BIBdecl}{\relax}
\BIBdecl

\end{thebibliography}


\begin{thebibliography}{15}                                                                


\bibitem{Childers} Krishnamurthy, A. and  Childers, D.,
``Two-channel speech analysis'', IEEE trans. on Acoustics, Speech and Signal Processing, 34:4, pp. 730-743, 1986.

\bibitem{PSOLA} Moulines, E. and Charpentier, F.,
``Pitch-synchronous waveform processing techniques for text-to-speech synthesis using diphones'', Speech Communication, vol. 9, pp. 453-467, 1990.


\bibitem{Stylianou} Stylianou, Y.,
``Removing linear phase mismatches in concatenative speech synthesis'', IEEE trans. on Speech and Audio Processing, vol. 9, issue 3, pp. 232-239, 2001.

\bibitem{VoiceConversion} Rentzos, D., Vaseghi, S., Turajlic, E., Qin Yan and Ching-Hsiang Ho,
``Transformation of speaker characteristics for voice conversion'', IEEE Workshop on Automatic Speech Recognition and Understanding, pp. 706-711, 2003.

\bibitem{Naylor} Gaubitch, N. and Naylor, P.,
``Spatio-temporal Averaging method for Enhancement of Reverberant Speech'', 15th Int. Conf. on Digital Signal Processing, pp. 607-610, 2007.

\bibitem{SpeakerIdentification} Gudnason, J. and Brookes, M.,
``Voice source cepstrum coefficients for speaker identification'', IEEE Int. Conf. on Acoustics, Speech and Signal Processing, pp. 4821-4824, 2008.

\bibitem{ZZT} Bozkurt, B., Couvreur, L. and Dutoit, T.,
``Chirp group delay analysis of speech signals'', Speech Comm., vol. 49, issue 3, pp. 159-176, 2007.

\bibitem{PSCELP} Guerchi, D. and Mermelstein, P.,
``Low-rate quantization of spectral information in a 4 kb/s pitch-synchronous CELP coder'', IEEE Workshop on speech coding, pp. 111–113, 2000.


\bibitem{Strube} Strube, H.W.,
``Determination of the instant of glottal closures from the speech wave'', JASA, vol. 56, pp. 1625-1629, 1974.

\bibitem{Yegna} Ananthapasmanabha, T. and Yegnanarayana, B.,
``Epoch extraction from linear prediction residual for identification of closed glottis interval'', IEEE Trans. Acoust., Speech and Signal Processing, vol. 27, no. 4, pp. 309-319, 1979.

\bibitem{Frobenius} Ma, Y., and Willems, L.,
``A Frobenius norm approach to glottal closure detection from the speech signal'', IEEE Trans. Speech Audio Processing, vol. 2, pp. 258-265, 1994.

\bibitem{Schnell}Schnell, K.,
``Estimation of Glottal Closure Instances from Speech Signals by Weighted Nonlinear Prediction'', Lecture Notes in Computer Science, Springer, pp. 221-229, 2007.

\bibitem{wavelet} Tuan, V. and d'Alessandro, C.,
``Robust glottal closure detection using the wavelet transform'', Proc. of the European Conference on Speech Technology, pp. 805-808, 1999.

\bibitem{GD} Smits, R. and Yegnanarayana, B.,
``Determination of instants of significant excitation in speech using group delay function'', IEEE Trans. Speech Audio Processing, vol. 3, no. 5, pp. 325-333, 1995.

\bibitem{Dypsa} Naylor, P., Kounoudes, A., Gudnason, J. and Brookes, M.,
``Estimation of glottal closure instants in voiced speech using the DYPSA algorithm'', IEEE Trans. Audio Speech Lang. Processing, vol. 15, no. 1, pp. 34-43, 2007.

\bibitem{Kahawara} Kawahara, H., Atake, Y. and Zolfaghari, P.,
``Accurate vocal event detection method based on a fixedpoint analysis of mapping from time to weighted average group delay'', Proc. ICSLP, pp. 664-667, 2000.

\bibitem{Murty} Murty, K. and Yegnanarayana, B.,
``Epoch Extraction From Speech Signals'', IEEE Trans. Audio Speech Lang. Processing, vol. 16, pp. 1602-1613, 2008.

\bibitem{Tunisian} Bouzid, A. and Ellouze, N.,
``Glottal Opening Instant Detection from Speech Signals'', Proc. of the 12th European Signal Processing Conference, 2004.

\bibitem{CMU} \emph{[Online]}, CMU ARCTIC speech synthesis databases, \emph{http://festvox.org/cmu\_arctic/}

\bibitem{DypsaWeb} \emph{[Online]}, Voicebox: Speech Processing Toolbox for Matlab, \emph{http://www.ee.ic.ac.uk/hp/staff/dmb/voicebox}











\end{thebibliography}


\end{document}